\documentstyle[twoside,fleqn,espcrc2,epsfig]{article}
\input {epsf.sty}
\input {psfig.sty}

\newcommand{\AmS}{{\protect\the\textfont2
  A\kern-.1667em\lower.5ex\hbox{M}\kern-.125emS}}

\title{Log Triviality in the Nambu - Jona-Lasinio Model}

\author{Simon Hands\address{Department of Physics,
        University of Wales Swansea, \\ 
        Singleton Park, Swansea SA2 8PP, United Kingdom}%
        \thanks{PPARC Advanced Fellow}
        }
       
\begin{document}

\begin{abstract}
Results are presented from a Monte Carlo simulation of the Nambu --
Jona-Lasinio model with SU(2)$\otimes$SU(2) chiral symmetry and
$N_f=2$ flavors of fermion. We show that fits to the equation of state
are sensitive to the shape and extent of the assumed scaling region;
the best fits favour a triviality scenario predicted by the large-$N_f$
approximation, in which the scalar degrees of freedom are fermion --
anti-fermion composites with wavefunction renormalisation constant 
vanishing logarithmically in the continuum limit. 
\end{abstract}

\maketitle

In this talk I report on work done with John Kogut; a more detailed
presentation has appeared~\cite{HK}. We have studied the critical
behaviour of the four-dimensional Nambu - Jona-Lasinio (NJL) model,
whose continuum Lagrangian density is given by
\begin{eqnarray}
{\cal L} &=& \bar\psi_i({\partial\!\!\! /}\,+m)\psi_i-
{g^2\over{2N_f}}\left[(\bar\psi_i\psi_i)^2-(\bar\psi_i\gamma_5\vec\tau
\psi_i)^2\right]\nonumber\\
&=&\bar\psi_i({\partial\!\!\! /}\,+m+\sigma+i\vec\pi.\vec\tau\gamma_5)
\psi_i+{{N_f}\over g^2}\mbox{tr}\Phi^\dagger\Phi,
\label{eq:Lag}
\end{eqnarray}
where $i$ runs over $N_f$ distinct fermion flavors, and in the second
line we introduce an auxiliary scalar field
$\Phi=\sigma+i\vec\pi.\vec\tau$ proportional to an SU(2) matrix.
The Lagrangian (\ref{eq:Lag}) has an $\mbox{SU(2)}_L\otimes\mbox{SU(2)}_R$
chiral symmetry $\psi_L\mapsto U\psi_L$, $\psi_R\mapsto V\psi_R$,
$\Phi\mapsto V\Phi U^\dagger$, which is broken explicity by the bare
fermion mass, and spontaneously by the development of a condensate
$\Sigma\equiv\langle\sigma\rangle\not=0$.

The NJL model has been proposed to describe both strong~\cite{NJL}
and weak~\cite{Bard} interaction physics; in the latter case the
Higgs is a composite $f\bar f$ state. Both the NJL model and the
conventional model of the Higgs sector, namely the O(4) linear sigma
model, are believed trivial in the sense that neither has an interacting
continuum limit. It has been shown that the lattice-regularised models
have equivalent predictive power for electroweak
phenomenology~\cite{Has}.

In this study we aim to characterise the triviality of the NJL model 
by fitting numerical data for $\Sigma$ to the model equation of state
(EOS)
\begin{equation}
m=A\left({1\over g^2}-{1\over
g_c^2}\right){\Sigma\over{\vert\ln\Sigma\vert^{q_1}}}+B
{\Sigma^3\over{\vert\ln\Sigma\vert^{q_2}}}.
\label{eq:eos}
\end{equation}
The logarithms are corrections to the EOS predicted by mean field
theory. Their presence signals triviality: the interacting theory 
becomes ill-defined in the continuum limit $\Sigma\to0$ (in lattice
units). Eq. (\ref{eq:eos}) has wide applicability; for the O($N$) sigma
model the exponents $q_{1,2}$ can be calculated using RG-improved
perturbation theory~\cite{Zinn}. For the NJL model, they can be estimated to
leading order in the $1/N_f$ expansion~\cite{Kocic}, 
and are qualitatively distinct
(see table). We wish to understand whether this difference is real
or merely an artifact of the approximation.

We have simulated a discretised version of (\ref{eq:Lag}) on a
$16^4$ lattice for $N_f=2$~\cite{HK}. Since we use staggered fermions,
this necessitates a fractional number of lattice flavors, and hence the
use of a hybrid molecular dynamics algorithm. We have accumulated a
high statistics dataset with bare mass values in the range
$m=0.05,\ldots,0.0025$. We also performed simulations directly in the
chiral limit $m=0$. In this case $\Sigma$ vanishes in a finite volume,
so instead we measured the mean field $\vert\Phi\vert$ defined via
\begin{equation}
\vert\Phi\vert=\left\langle\sqrt{{\textstyle{1\over2}}\mbox{tr}
\bar\Phi^\dagger\bar\Phi}\right\rangle\;\;\;;\;\;\;\bar\Phi={1\over V}\sum_x
\Phi(x).
\end{equation}

\begin{figure}[htb]
\centerline{
\setlength\epsfxsize{230pt}
\epsfbox{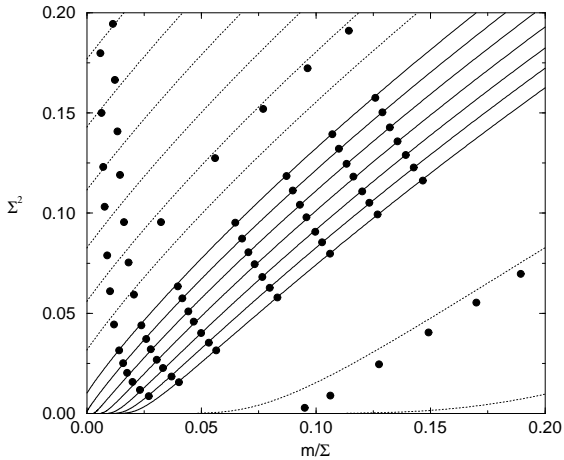}
}
\caption{Fisher plot of the ferromagnetic fit.}
\label{fig:ferromagnet}
\end{figure}

\begin{figure}[htb]
\centerline{
\setlength\epsfxsize{230pt}
\epsfbox{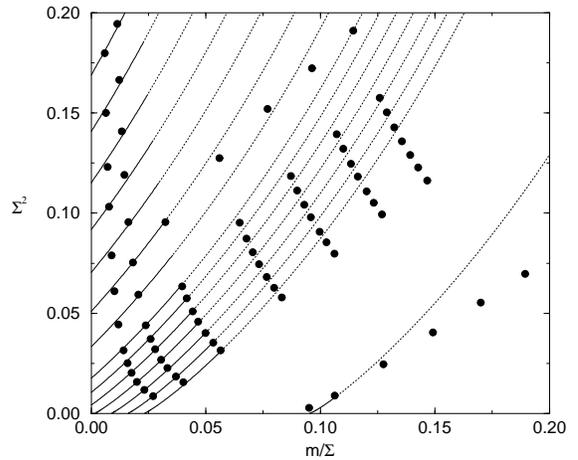}
}
\caption{Fisher plot of the composite fit.}
\label{fig:composite}
\end{figure}

\begin{table*}[hbt]
\setlength{\tabcolsep}{1.5pc}
\newlength{\digitwidth} \settowidth{\digitwidth}{\rm 0}
\catcode`?=\active \def?{\kern\digitwidth}
\caption{Values for the correction exponents}
\begin{tabular*}{\textwidth}{@{}l@{\extracolsep{\fill}}rrrr}
\hline
                 & \multicolumn{2}{l}{Analytical predictions} 
                 & \multicolumn{2}{l}{Numerical fits} \\
\cline{2-3} \cline{4-5}
                 & \multicolumn{1}{r}{O(4) sigma model} 
                 & \multicolumn{1}{r}{large-$N_f$ NJL} 
                 & \multicolumn{1}{r}{ferromagnetic} 
                 & \multicolumn{1}{r}{composite}         \\
\hline
$q_1$            &  0.5 & 0.0 & 0.786(19) & 0.016(11) \\
$q_2$            &  1.0 & -1.0 & 0.372(17)& -0.553(18) \\
$\chi^2$/dof     &      &      & 2.5     &  4.6 \\
\hline
\end{tabular*}
\end{table*}

In order to find a satisfactory fit to (\ref{eq:eos}) it is 
necessary to make some assumptions about the size and shape of the
scaling window about the critical point, and hence which data 
to include. We have found that our results for $q_i$ depend very
sensitively on the window chosen. For instance, if we include all 
mass values, but exclude couplings outside the range [0.52,0.55]
then we obtain values, shown in the table, which qualitatively 
resemble the analytic prediction for the O(4) sigma model: we label
this fit ``ferromagnetic''. If, on the other hand, we include all
coupling values but exclude masses greater than 0.01, then the fit
changes, the value of $q_1$ becoming almost zero, and $q_2$
becoming negative. We refer to this fit as ``composite'' because
it agrees with the $1/N_f$ treatment of the NJL model. The results
of the fits are displayed in Figs.~\ref{fig:ferromagnet},
\ref{fig:composite} in the form of a Fisher
plot of $\Sigma^2$ vs. $m/\Sigma$. This would yield straight lines of
constant coupling as $m\to0$ for a mean field EOS.

It would appear that logarithmic corrections are hard to pin down from
EOS fits alone. Both fits have sufficiently large $\chi^2$ to
be questionable. Fortunately there is an an alternative probe,
involving data for $\vert\Phi\vert$ taken at $m=0$. On a finite system
we expect $\vert\Phi\vert$ to exceed $\Sigma_0$, the true
order parameter obtained by extrapolating the EOS to the chiral limit,
because $\vert\Phi\vert$ also receives contributions from fluctuations
of Goldstone modes, which average to zero only in the thermodynamic
limit.
\begin{figure}[htb]
\centerline{
\setlength\epsfxsize{230pt}
\epsfbox{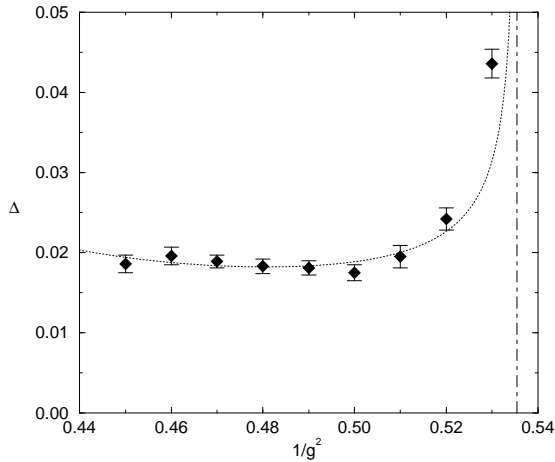}
}
\caption{$\Delta$ vs. $1/g^2$: the vertical line shows the fit for
$1/g_c^2$}
\label{fig:delta}
\end{figure}
The difference $\Delta=\vert\Phi\vert-\Sigma_0$ can be
calculated using chiral perturbation theory to be of the
form~\cite{Gock}
\begin{equation}
\Delta\propto{Z\over{\Sigma_0 L^2}},
\label{eq:delta}
\end{equation}
where $Z$ is the wavefunction renormalisation constant, defined as the
coefficient of $1/p^2$ in the pion propagator in the IR limit. The
values of $\Delta$ obtained using the chiral extrapolation of the
composite EOS fit obtained by setting 
$q_1=0$, $q_2=-1$~\cite{HK} are shown as a function of $1/g^2$ 
in Fig.~\ref{fig:delta} ($\Delta$ extracted using the ferromagnetic
EOS changes sign over the range and makes no physical sense).

At first sight the data look incompatible with (\ref{eq:delta}),
since $\Delta$ is approximately constant for $1/g^2\in[0.45,0.52]$,
while $\Sigma_0$ falls by a factor of 3 in the same range.
However, we must take into account the dependence of $Z$ on
$\Sigma_0$. For models such as the sigma model in which the Higgs
is elementary, $Z$ is perturbatively close to 1 approaching the
continuum limit, and hence $\Delta\propto1/\Sigma_0$. For a composite
Higgs, however, we expect $Z$ to vanish logarithmically in the same
limit~\cite{Kocic}\cite{Eguchi}, 
and hence $\Delta\propto1/\Sigma_0\vert\ln\Sigma_0\vert$.
The latter form yields a reasonable fit as shown in 
Fig.~\ref{fig:delta}.

To conclude, this independent study of the finite volume correction
seems to prefer the composite EOS predicted in the large-$N_f$
approach. It remains an interesting open question whether this result
is generic to four-dimensional models in which scalars are composite, 
or whether other universality classes exist defined by alternative
lattice formulations~\cite{AliKhan}.

\end{document}